\pdfoutput=1
\documentclass{aa}  
\usepackage{graphicx}
\usepackage{txfonts}
\usepackage{natbib}
\bibpunct{(}{)}{;}{a}{}{,}
\newcommand{\jdu}{\mbox{$J$=2$-$1}}
\newcommand{\juc}{\mbox{$J$=1$-$0}}
\newcommand{\kms}{\mbox{~km\,s$^{-1}$}}
\begin{document}
   \title{SiO masers from AGB stars in the vibrationally excited $v$=1,
$v$=2, and $v$=3 states
}

\titlerunning{}

   \author{J.-F.\ Desmurs\inst{1} \and 
          V. Bujarrabal\inst{2} \and M.\ Lindqvist\inst{3} \and
          J.\ Alcolea\inst{1} \and R.\ Soria-Ruiz\inst{1} \and
          P.\ Bergman\inst{3} 
 }

   \offprints{J.-F. Desmurs}

   \institute{
             Observatorio Astron\'omico Nacional (OAN-IGN),
             C/ Alfonso XII, 3, E-28014 Madrid, Spain\\
              \email{jf.desmurs@oan.es}
\and
             Observatorio Astron\'omico Nacional (OAN-IGN),
             Apartado 112, E-28803 Alcal\'a de Henares, Spain
\and
             Onsala Space Observatory, Dept. of Earth and Space
             Sciences, Chalmers University of Technology, 43992
             Onsala, Sweden
           }
\date{Submitted: 31th January 2014; accepted: 17th March 2014 }
\abstract 
{The $v$=1 and $v$=2 \juc\ (43 GHz), and $v$=1 \jdu\ (86 GHz) SiO
  masers are intense in Asymptotic Giant Branch (AGB) stars and have
  been mapped using Very Long Baseline Interferometry (VLBI) showing
  ring-like distributions. Those of the $v$=1, $v$=2 \juc\ masers are
  similar, but the spots are rarely coincident, while the $v$=1 \jdu\
  maser arises from a well separated region farther out. These relative
  locations can be explained by models tools that include the overlap
  of two IR lines of SiO and H$_2$O. The $v$=3 \juc\ line is not
  directly affected by any line overlap and its spot structure and
  position, relative to the other lines, is a good test to the standard
  pumping models.  }
{The aim of this project are to gain insight into the properties and
  the general theoretical considerations of the different SiO masers
  that can help to understand them.}
{We present single-dish and simultaneous VLBI observations of the
  $v$=1, $v$=2, and $v$=3 \juc\ maser transitions of $^{28}$SiO in
  several AGB stars. The results are compared to the predictions of
  radiative models of SiO masers that both include and not include the
  effect of IR line overlap. }
{The spatial distribution of the SiO maser emission in the $v$=3 \juc\
  transition from AGB stars is systematically composed of a series of
  spots that occupy a ring-like structure (as often found in SiO
  masers). The overall ring structure is extremely similar to that
  found in the other 43 GHz transitions and is very different from the
  structure of the $v$=1 \jdu\ maser. The positions of the individual
  spots of the different 43 GHz lines are, however, very rarely
  coincident, which in general is separated by about 0.3 AU (between 1
  and 5 mas). These results are very difficult to reconcile with
  standard pumping models, which predict that the masers of rotational
  transitions within a given vibrational state require very similar
  excitation conditions (since the levels are placed practically at the
  same energy from the ground), while the transitions of different
  vibrational states (which are separated by an energy of 1800 K)
  should appear in different positions. However, models including line
  overlap tend to predict $v$=1, $v$=2, $v$=3 \juc\ population
  inversion to occur under very similar conditions, while the
  requirements for $v$=1 \jdu\ appear clearly different, and are
  compatible with the observational results.  }
{}
\maketitle

\section{Introduction}

Many Asymptotic Giant Branch (AGB) stars have been mapped in SiO maser
emission in the \juc\ $v$=1 and $v$=2 lines\footnote{In this paper,
$v$=1, $v$=2, $v$=3 refers to masers at the $v$=1, $v$=2, or $v$=3
states.}  using Very Long Baseline Interferometry (VLBI)
\citep[][etc]{diam94, desmurs00, cot06}. The maser emission is found to
form a ring of spots at a few stellar radii from the center of the
star. In general, both distributions are similar, although the spots
are very rarely coincident, and the $v$=2 ring is slightly closer to
the star than the $v$=1 ring \citep[see e.g.][]{desmurs00}.

The similar distributions of the $v$=1, $v$=2 \juc\ transitions were
first interpreted as favoring collisional pumping, because the
radiative mechanism discriminates the location of the two masers
more. On the contrary, the lack of true coincidence was used to argue
in favor of radiative pumping, which leads to the well-known,
long-lasting discrepancy in the interpretation of the $v$=1, $v$=2
\juc\ maps in terms of pumping mechanisms; see detailed discussion in
\cite{desmurs00}.

Our understanding of this topic changed dramatically when the first
comparisons between the $v$=1 \juc\ and $v$=1 \jdu\ maser
distributions were performed \citep{soria04,soria05,soria07}. In
contrast to predictions from both models (radiative and collisional),
the $v$=1 \jdu\ maser spots are systematically found to occupy a ring
with a significantly larger radius (by about 30\%) than that of $v$=1
\juc, where both spot distributions being completely
unrelated. \cite{soria04} explained these unexpected results by
invoking line overlap between the ro-vibrational transitions $v$=2--1
\juc\ of SiO and $v_2$=0--1 $J_{K_a,K_c}$=$12_{7,5}$--$11_{6,6}$ of
H$_2$O. According to \cite{soria04}, this phenomenon, which was
first proposed by \cite{olof81} to explain the anomalous weakness of
the $v$=2 \jdu\ SiO maser, would also introduce a strong coupling of
the $v$=1 and $v$=2 \juc\ lines, explaining their similar
distributions.

In the simplest theoretical interpretation \citep[not including line
overlap, see][]{buj81,buj94,loc92,hum02}, the $v$=3 \juc\ emission
requires completely different excitation conditions than the other less
excited lines, since the $\Delta v$=1 energy separation is very high,
$\sim$1800~K. The $v$=3 \juc\ spatial distribution should in principle
be different from the $v$=1, $v$=2 \juc\ ones and, of course, the $v$=1
\jdu\ maser, and placed in a still more inner ring. However, we have
seen that line overlap strongly affects the $v$=1 and $v$=2 \juc\ maser
pumping. In particular, this phenomenon changes the conditions required
to pump both lines, which now tend to require higher densities.  In
this paper, we compare the $v$=1, $v$=2, $v$=3 \juc\ maser distribution
and analyzed the results in the framework of the pumping models.

\section{Observational data}

The main goal of this paper is to present maps with sub-arc resolution
in the $v$=3 \juc\ line and compare them to maps of $v$=1 and $v$=2
\juc\ lines.  $v$=1 and $v$=2 are usually strong masers easy to observe
in VLBI, but this is not the case of the much weaker $v$=3 lines. This
line can be sometimes very intense but it is strongly variable
\citep{alco89} both in time (with characteristic time scales of a few
months) and from object to object. To maximize the success of the
interferometric observations, we first initiated a single-dish
monitoring program of a sample of 19 AGB stars showing SiO masers.
Once an AGB star had a $v$=3 flux density above $\sim$5 Jy, we
triggered the Very Large Baseline Array (VLBA) to observe the star in
the weeks immediately following the single-dish detection.

\subsection{Single-dish monitoring of $v$=3 \juc\ emission} 
\begin{table}[h]
\centering
\normalsize
\begin{tabular}{|l|r|r|r|}
\hline
{\bf Source} &  RA(J2000) & Dec(J2000) & $V_{LSR}$\\
\hline
Y Cas    & 00:03:21.300& +55:40:50.00 & -18.0 \\
IRC+10011& 01:06:25.980& +12:35:53.00 & +11.0 \\ 
W And    & 02:17:32.961& +44:18:17.77 & -38.0 \\ 
o Cet    & 02:19:20.793& -02:58:39.51 & +45.0 \\  
IK Tau   & 03:53:28.840& +11:24:22.60 & +34.0 \\  
U Ori    & 05:55:49.169& +20:10:30.69 & -36.0 \\  
TX Cam   & 05:00:50.390& +56:10:52.60 & +09.0 \\  
V Cam    & 06:02:32.297& +74:30:27.10 & +08.0 \\  
R Cnc    & 08:16:33.828& +11:43:34.46 & +17.0 \\  
R Leo    & 09:47:33.490& +11:25:43.65 & -01.0 \\  
R LMi    & 09:45:34.283& +34:30:42.78 & +02.0 \\  
RU Her   & 16:10:14.515& +25:04:14.34 & -10.0 \\ 
U Her    & 16:25:47.471& +18:53:32.87 & -16.0 \\  
R Aql    & 19:06:22.252& +08:13:48.01 & +46.0 \\  
GY Aql   & 19:50:06.336& -07:36:52.45 & +33.0 \\  
$\chi$ Cyg& 19:50:33.922& +32:54:50.61 & +12.0 \\  
RR Aql   & 19:57:36.060& -01:53:11.33 & +30.0\\  
$\mu$ Cep& 21:43:30.461& +58:46:48.17 & +24.0 \\  
R Cas    & 23:58:24.873& +51:23:19.70 & +25.0 \\  
\hline
\end{tabular}
 \caption{List of sources monitored using the Onsala 20~m Telescope.}
\label{table-sou}
\end{table}
\noindent

We monitored a list of 19 AGB stars, as seen in Table \ref{table-sou},
using the 20-m antenna at the Onsala Space Observatory for three years
(from December 2008 to December 2011). We simultaneously observed the
$v$=1, $v$=2, $v$=3 \juc\ lines at 43 GHz in both right and left
circular polarizations (except for the two first runs, which are mostly
devoted to feasibility check and which we only observed $v$=1,
$v$=3). The spectral resolution was 25~kHz, equivalent to
$\sim$0.2~\kms.  As we were interested in searching for strong maser
emission that could be later mapped with the VLBA, each source was
observed with about 10 minutes of integration time in position
switching mode. For a typical $T_{\rm sys}$ of $\sim$150~K, this gave
us a rms noise of $\sim$0.1~K, which was good enough for our detection
purpose.
%
   \begin{figure}
   \centering
 \includegraphics[angle=-90,width=0.24\textwidth]{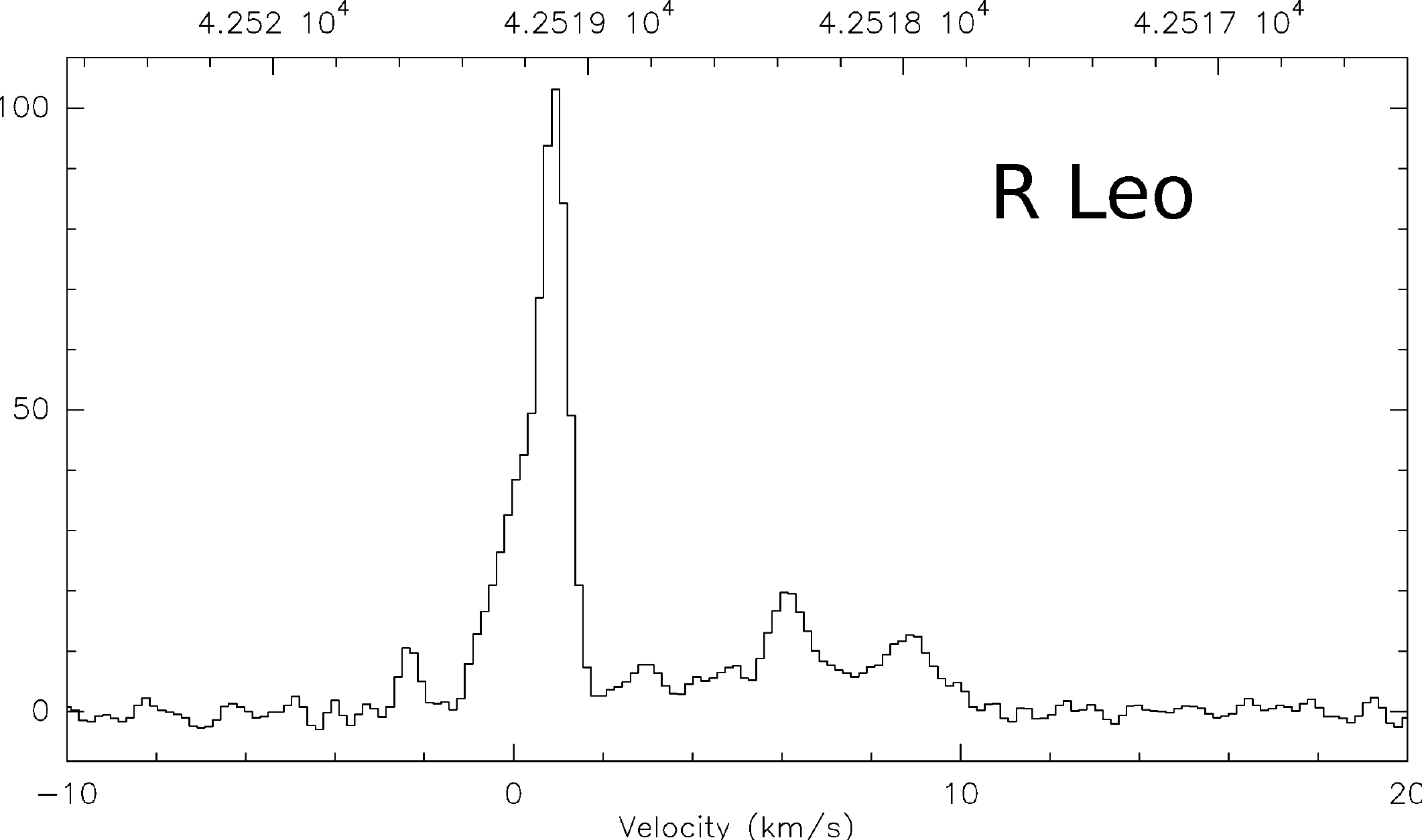}
 \includegraphics[angle=-90,width=0.24\textwidth]{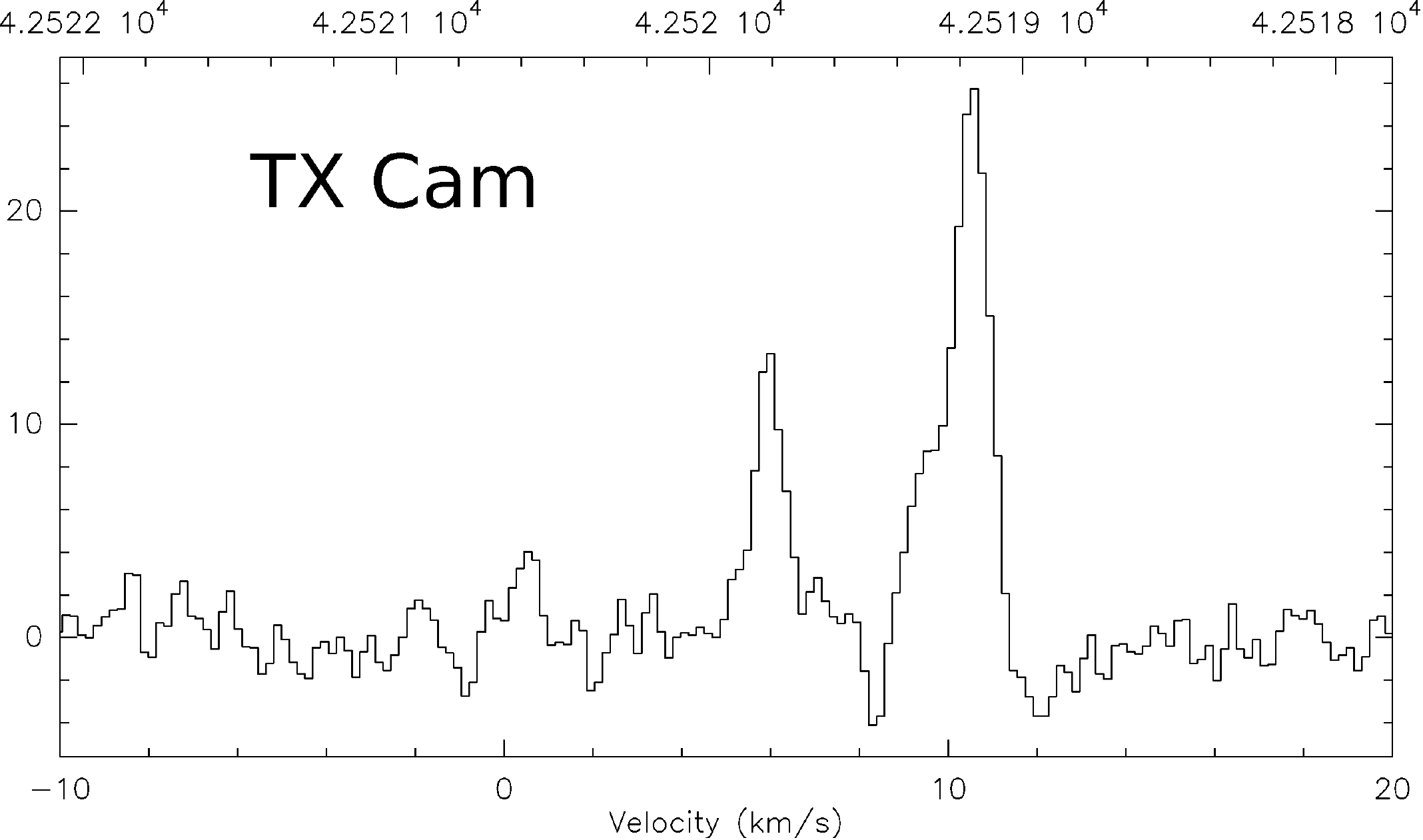}
 \includegraphics[angle=-90,width=0.24\textwidth]{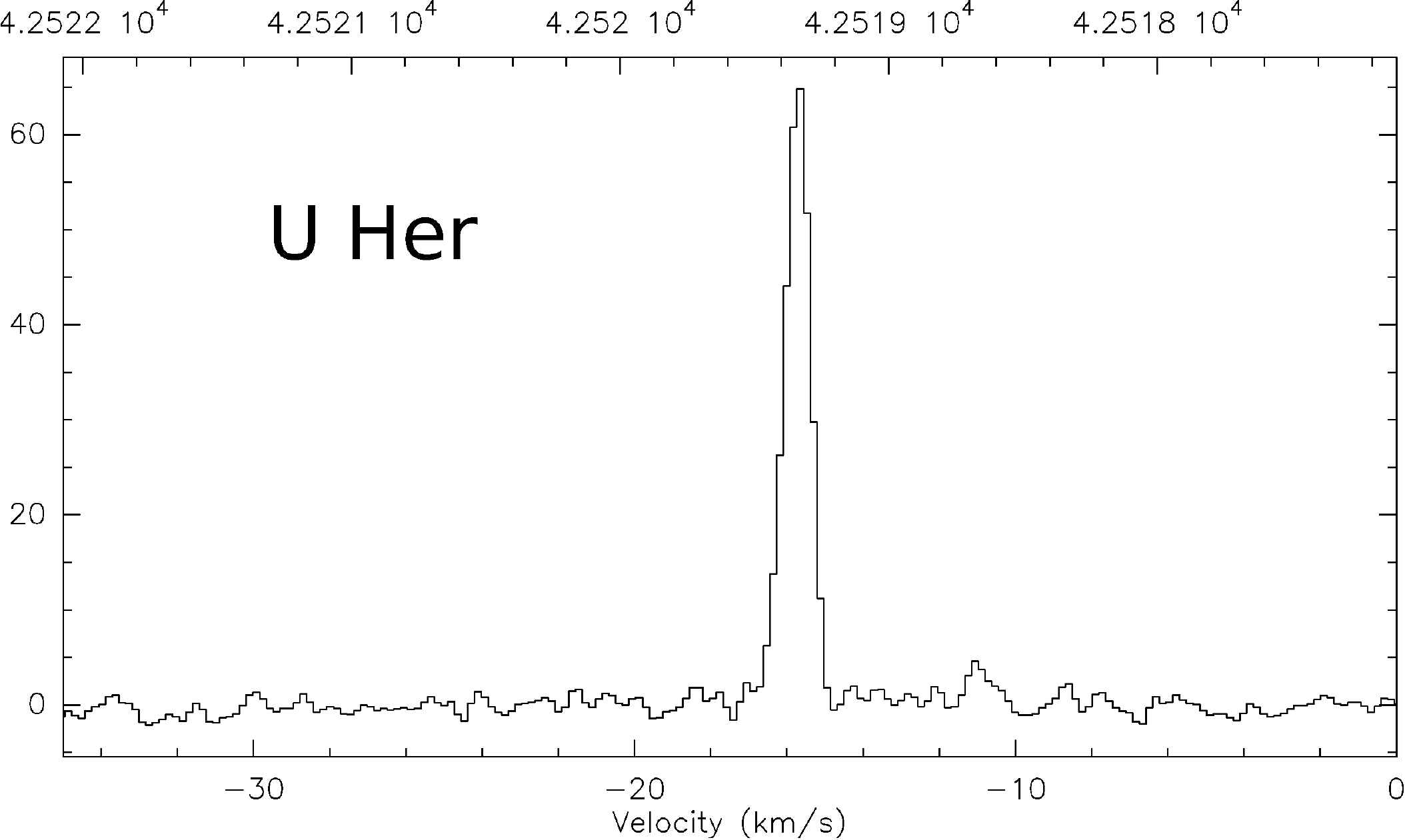}
 \includegraphics[angle=-90,width=0.24\textwidth]{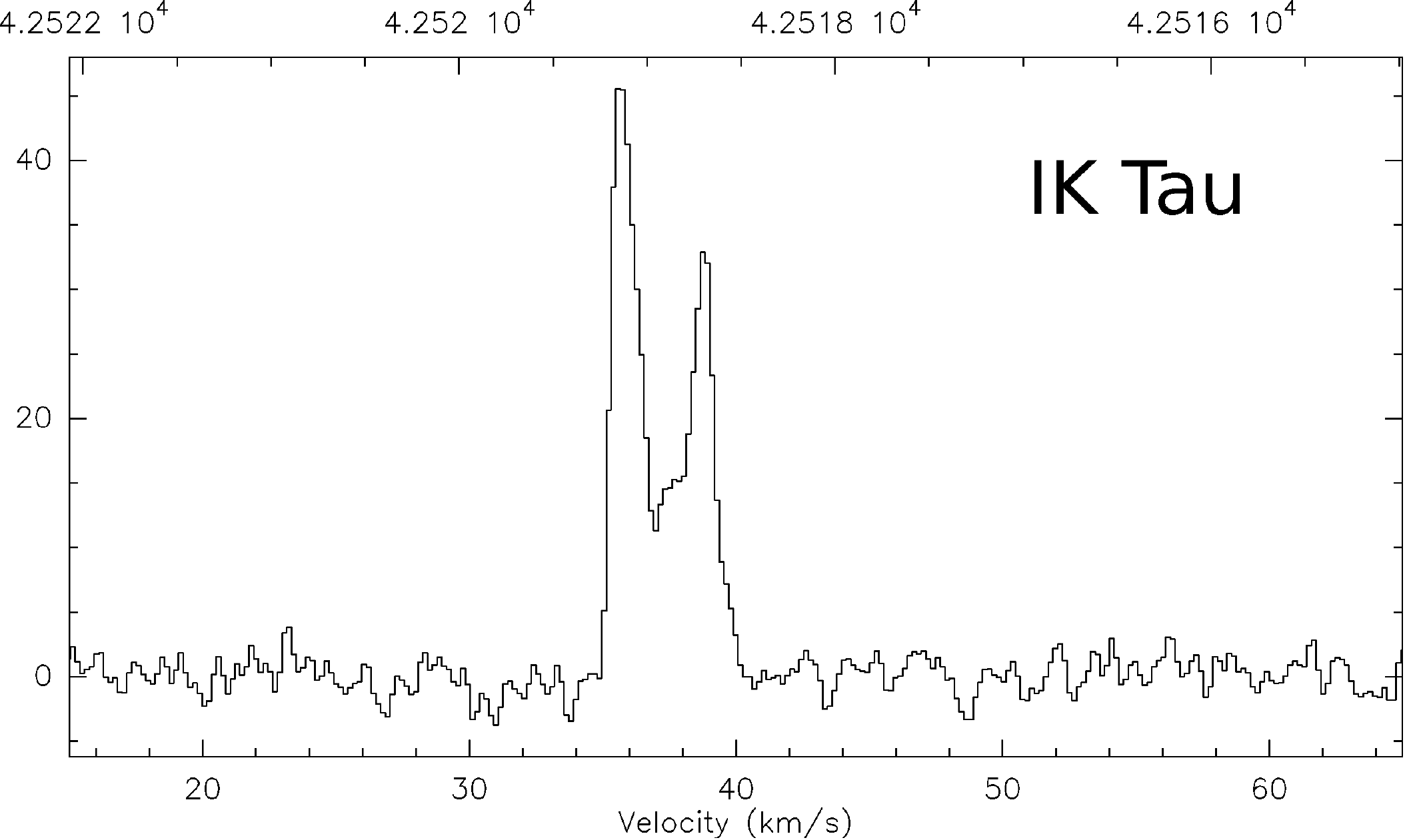}
 \caption[]{Single-dish spectra of the SiO \juc\ $v$=3 maser emission
   from the AGB stars, R~Leo (November, 2009), TX~Cam (January, 2010),
   U~Her (April, 2011), and IK~Tau (October, 2011). The intensity scale
   is in Jy.}
   \label{fig-spect}
   \end{figure}

\begin{figure*}
  \centering
   \includegraphics[angle=0,width=0.3\textwidth]{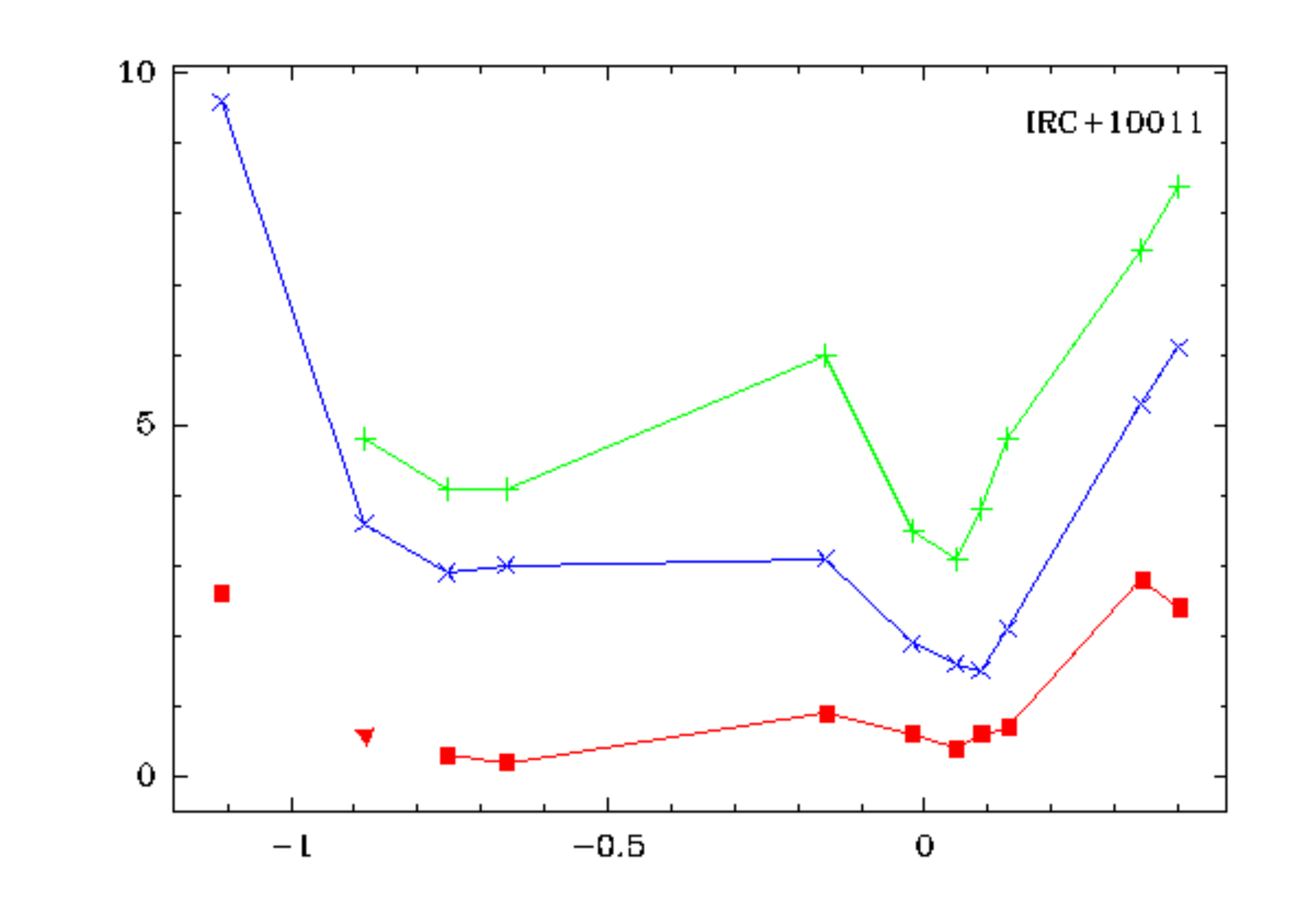}
   \includegraphics[angle=0,width=0.3\textwidth]{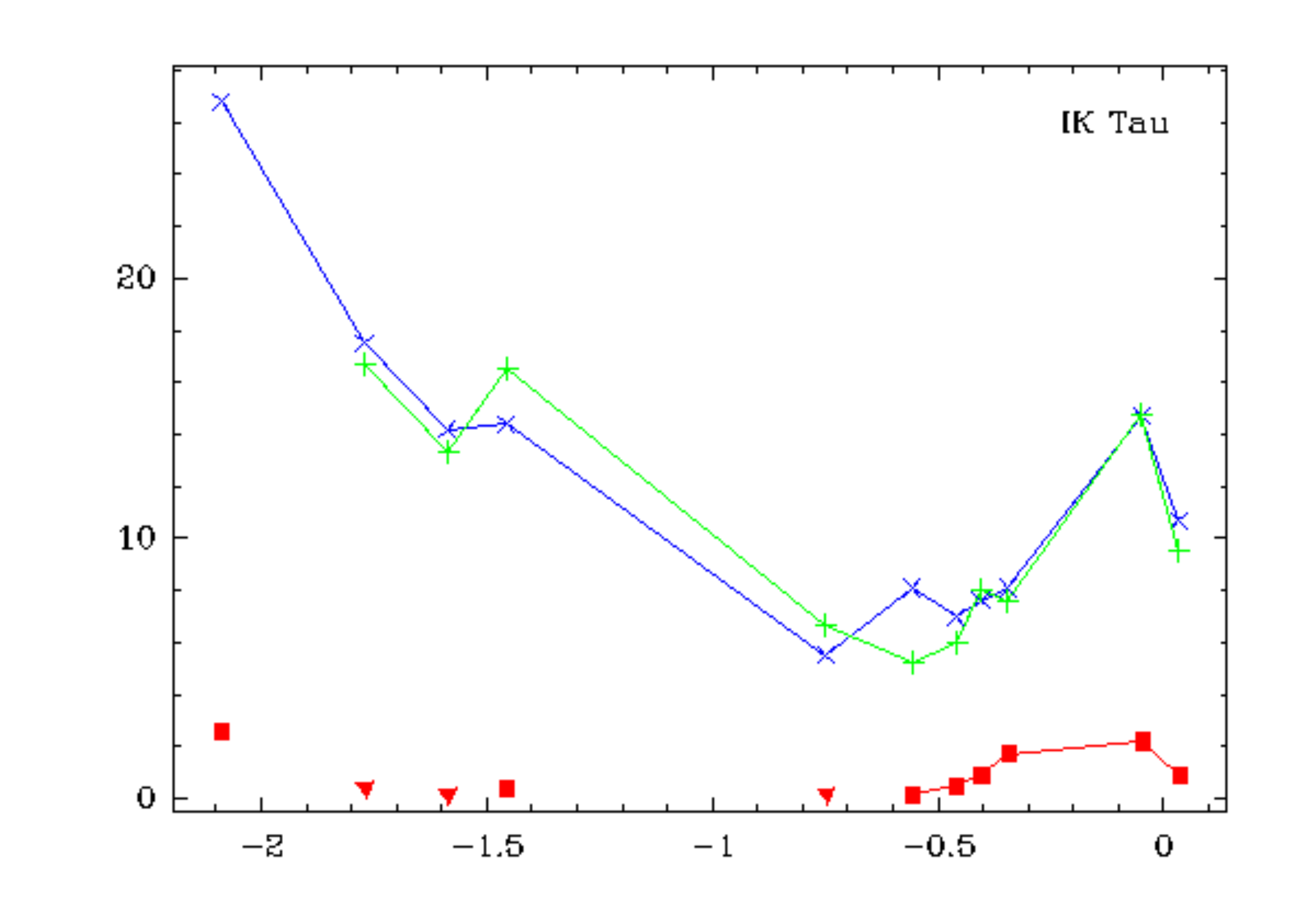}
   \includegraphics[angle=0,width=0.3\textwidth]{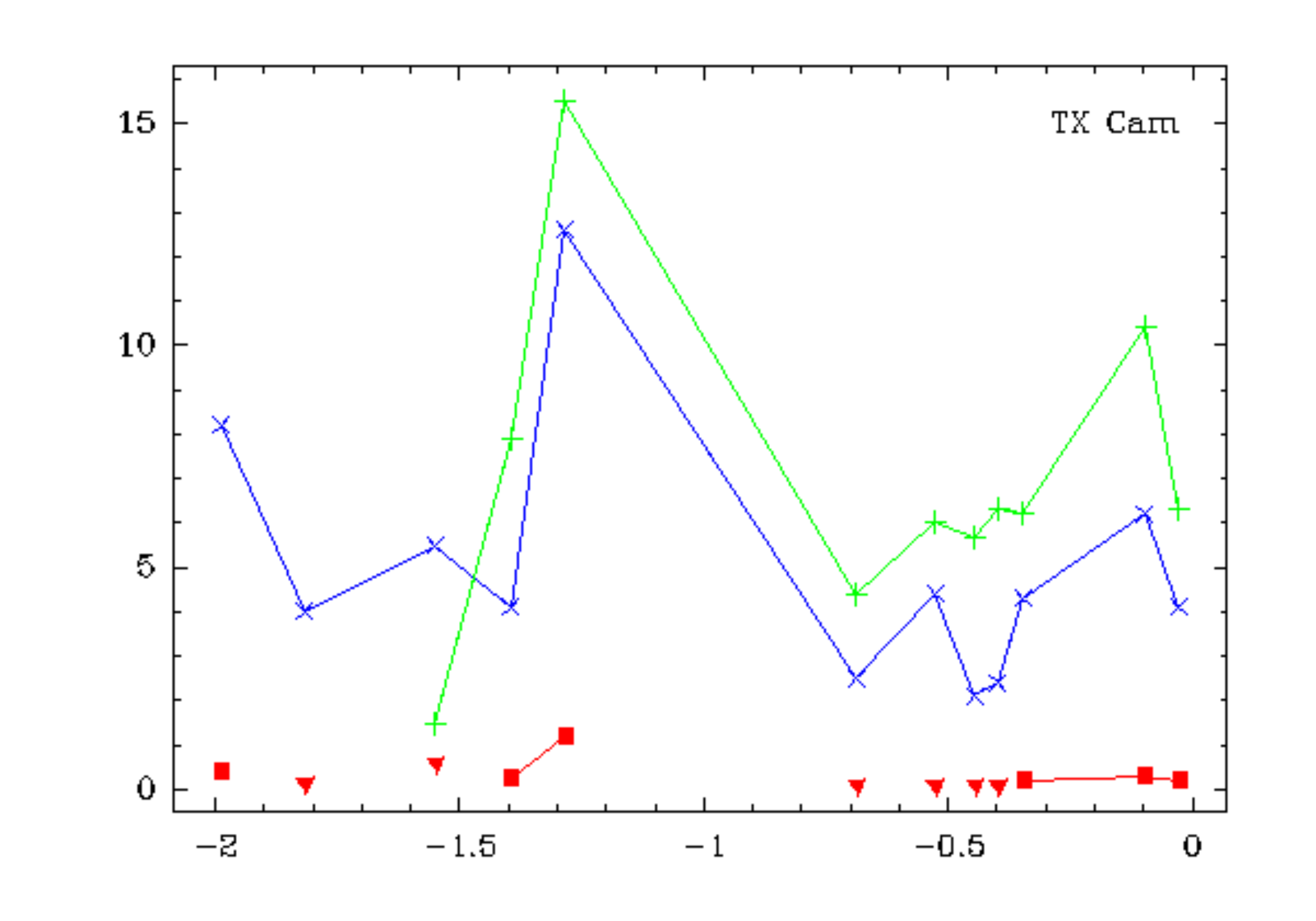}
   \includegraphics[angle=0,width=0.3\textwidth]{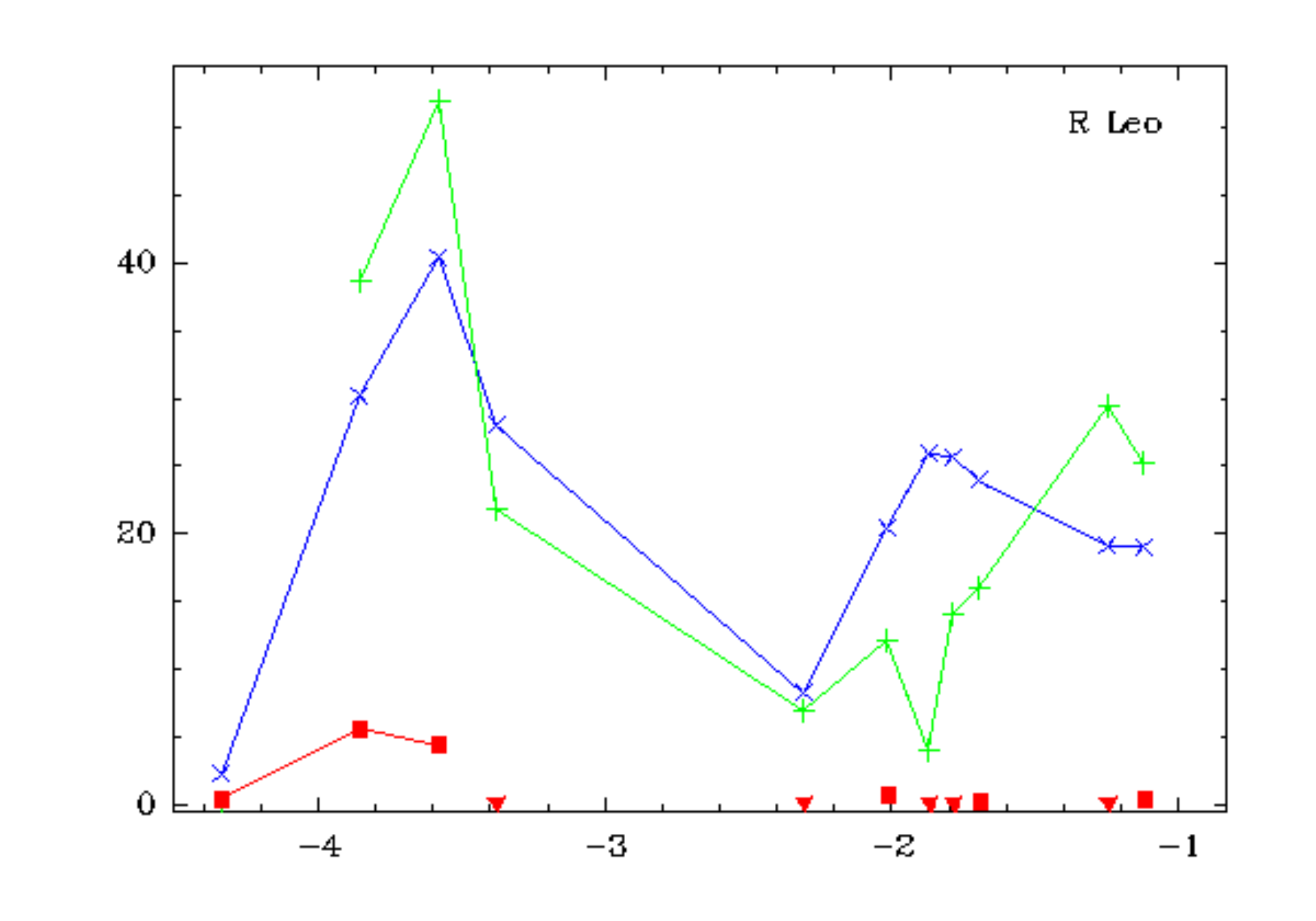}
   \includegraphics[angle=0,width=0.3\textwidth]{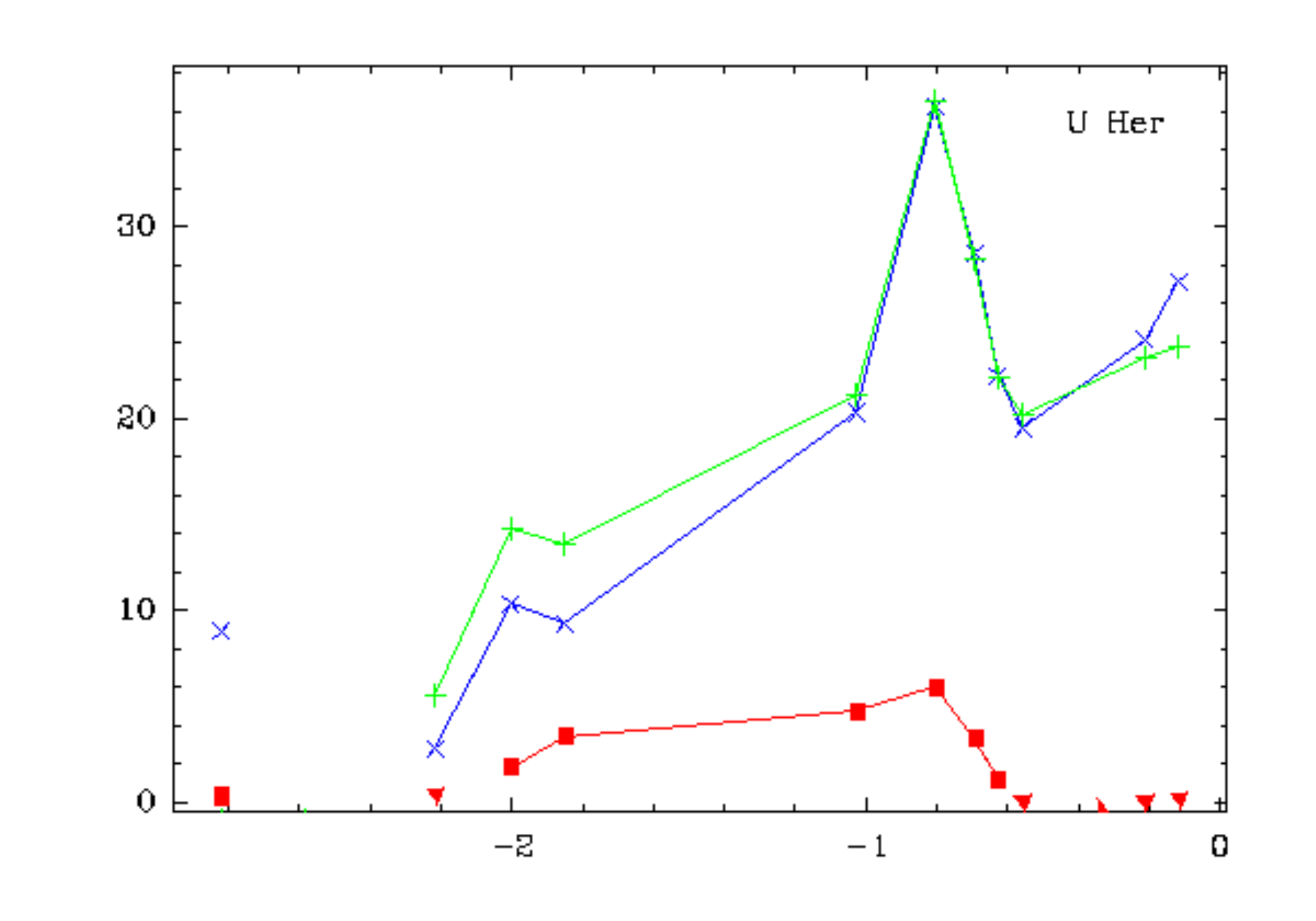}
   \includegraphics[angle=0,width=0.3\textwidth]{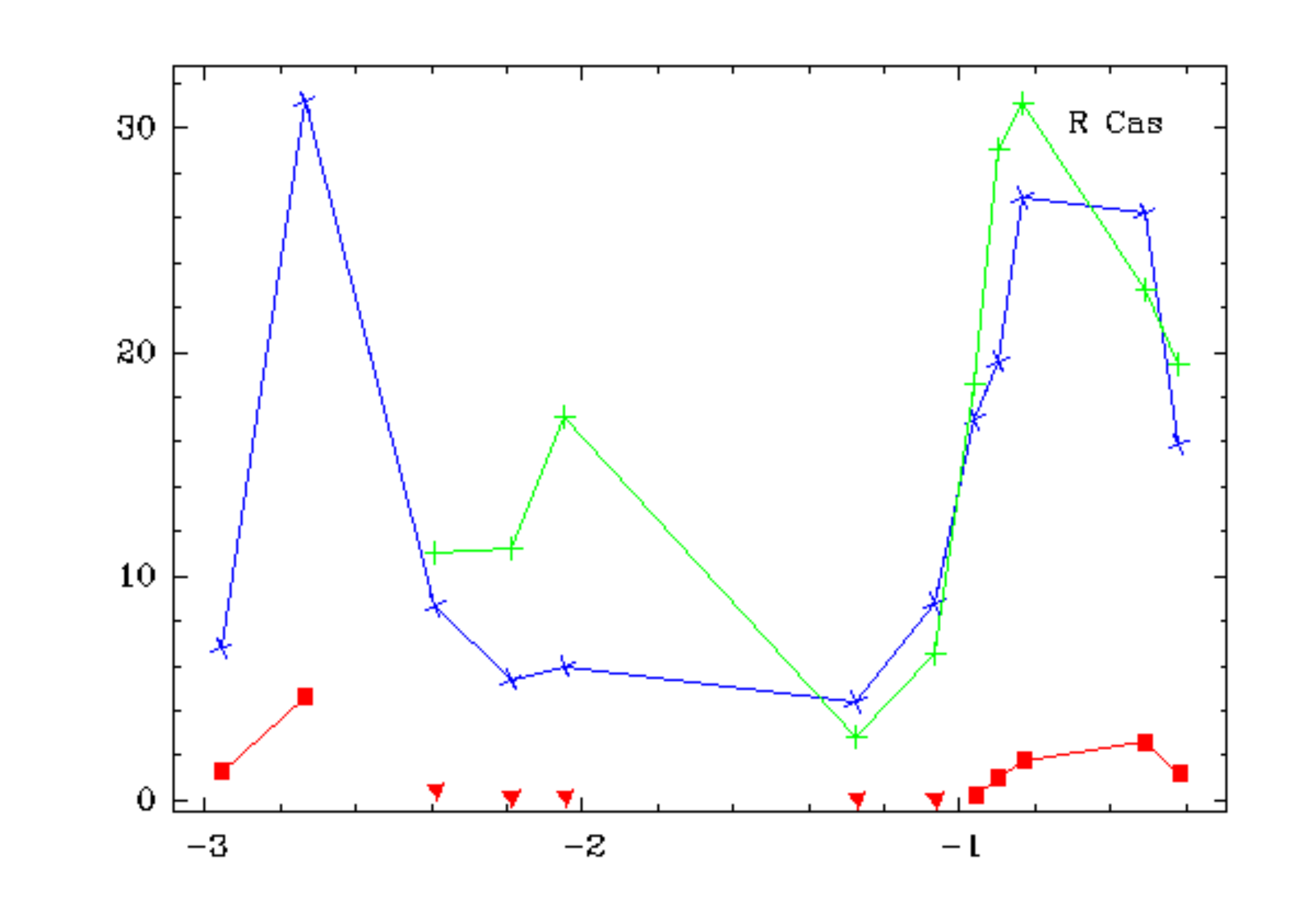}
   \includegraphics[angle=0,width=0.3\textwidth]{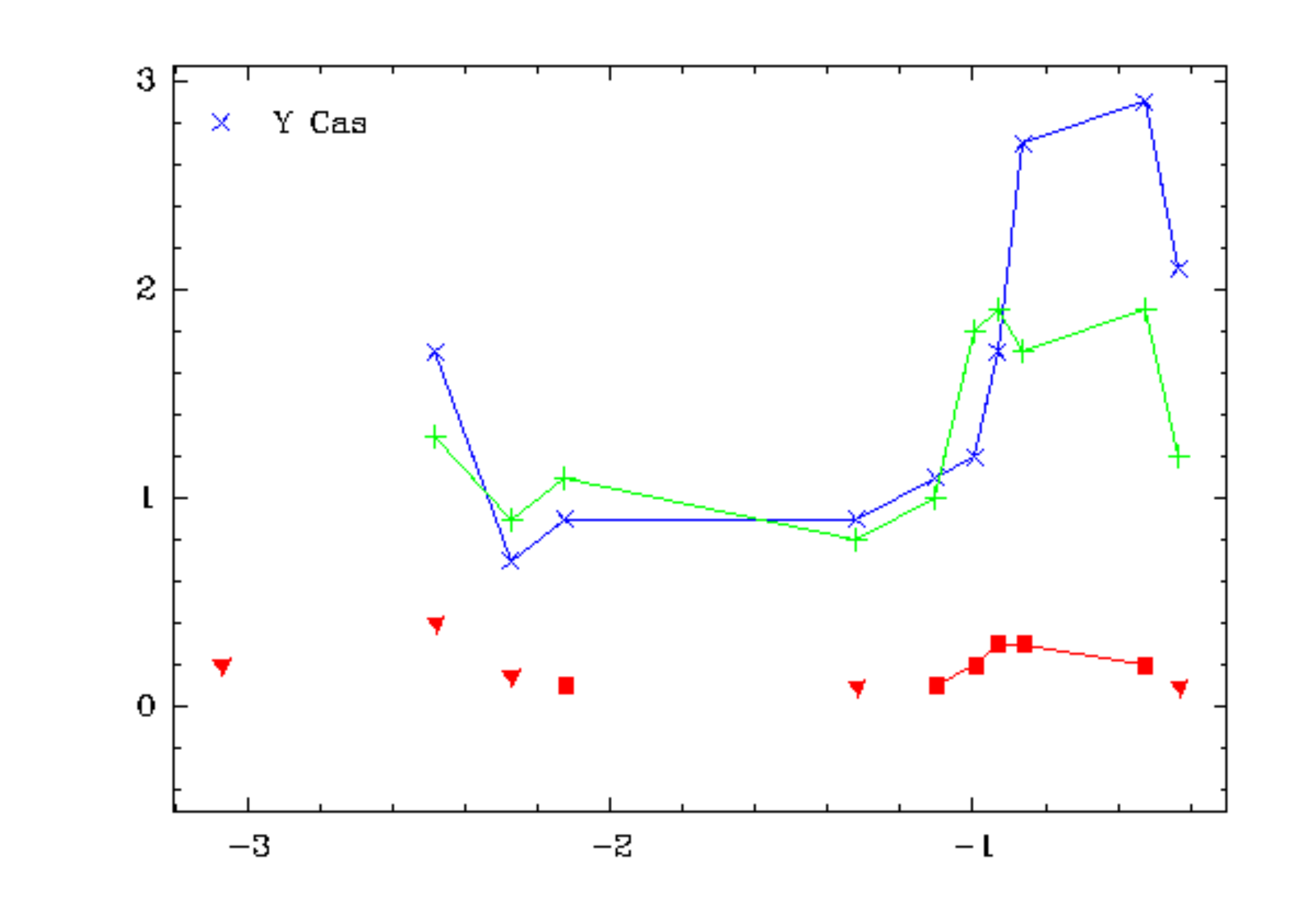}
  \caption[]{ Flux variation intensity (in Jy) with time (phase of the
  variability cycle of the source) from our monitoring with the Onsala
  telescope, of $v$=1, $v$=2, $v$=3 (respectively blue cross, green
  plus and red square, triangles denote upper limits detection) for the
  sources detected more often than in two epochs in $v$=3 \juc. See
  observing dates in Table \ref{table-moni}.}
  \label{fig-flux}
\end{figure*}

In total, we carried out 12 observing runs and we detected several
sources in $v$=3 \juc\ emission at different epochs; see Table
\ref{table-moni} for a summary of our monitoring results. For each
epoch, we searched for candidates that were good enough to be mapped
with the VLBA. In Fig. \ref{fig-spect}, we show the single dish $v$=3
\juc\ spectra in dates close to the VLBI observations of the four
trigger candidates that were selected: R~Leo (observing date November
07, 2009, optical phase $\phi=\sim$0.42), TX~Cam (observing date
January 08, 2010, optical phase $\phi=\sim$0.71), U~Her (observing date
April 22, 2011, optical phase $\phi=\sim$0.31), and IK~Tau (observing
date October 30, 2011, optical phase $\phi=\sim$0.95).  Intensity scale
is in Jy; the conversion factor used was F(Jy)/$T_a^*$(K) = 20.

In Fig.\ \ref{fig-flux}, we graphically present the flux peak intensity
as a function of the variability cycle for sources that were detected
at least a couple of times in $v$=3 during our monitoring with the
Onsala 20\,m telescope. From our 19 sources, seven fulfill this
condition: IRC+10011, IK~Tau, TX~Cam, R~Leo, U~Her, R~Cas, and Y~Cas.
As we can see in Fig.\ \ref{fig-flux}, the general trend for all three
lines is very similar, and $v$=3 seems to follow the same flux
variation as the other two lines but is an order of magnitude weaker.

\subsection{VLBA mapping}

   \begin{figure*}[h]
\begin{center}
 \includegraphics[angle=0,width=0.45\textwidth]{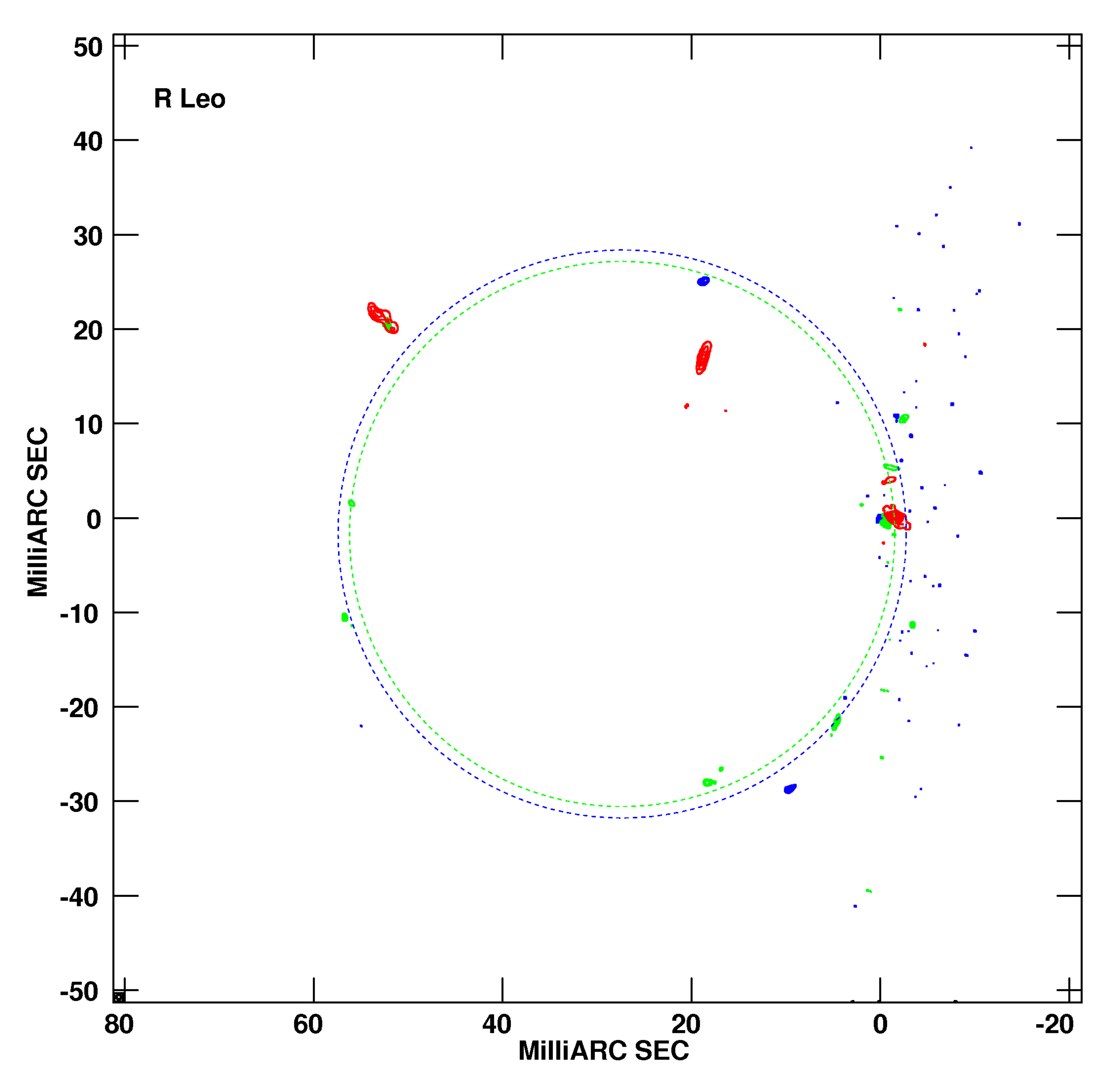}
 \includegraphics[angle=0,width=0.45\textwidth]{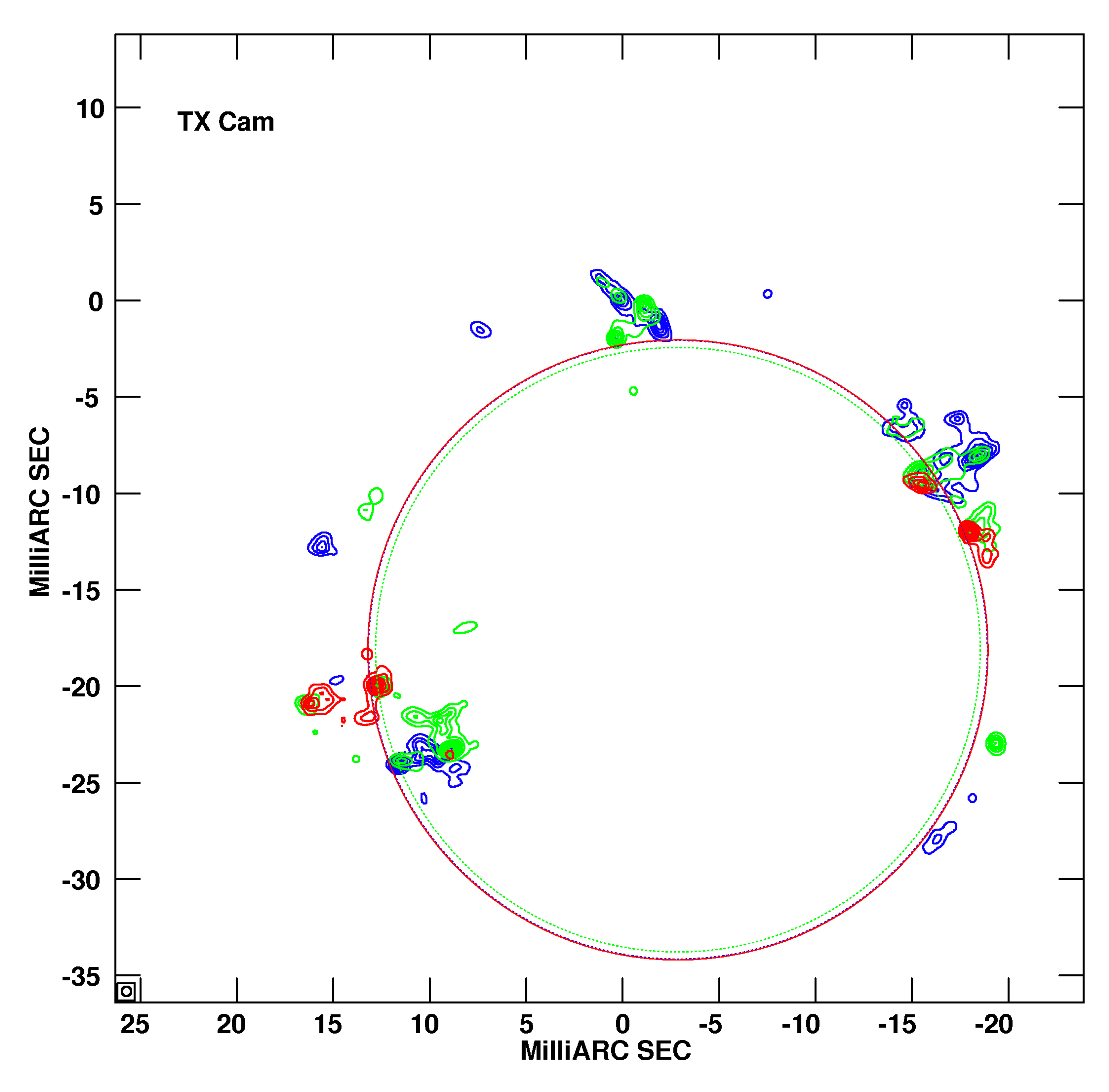}
 \includegraphics[angle=0,width=0.45\textwidth]{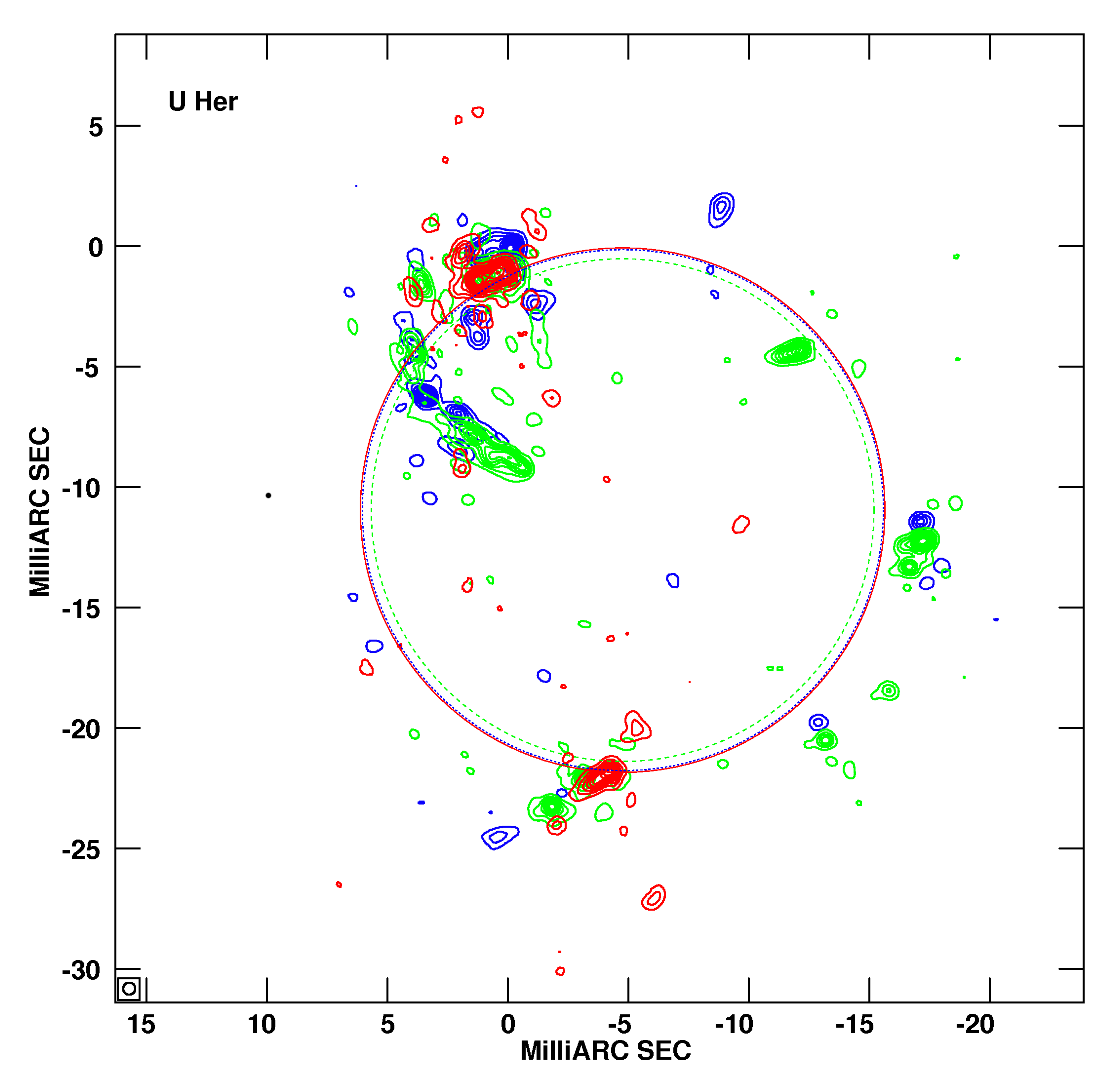}
 \includegraphics[angle=0,width=0.45\textwidth]{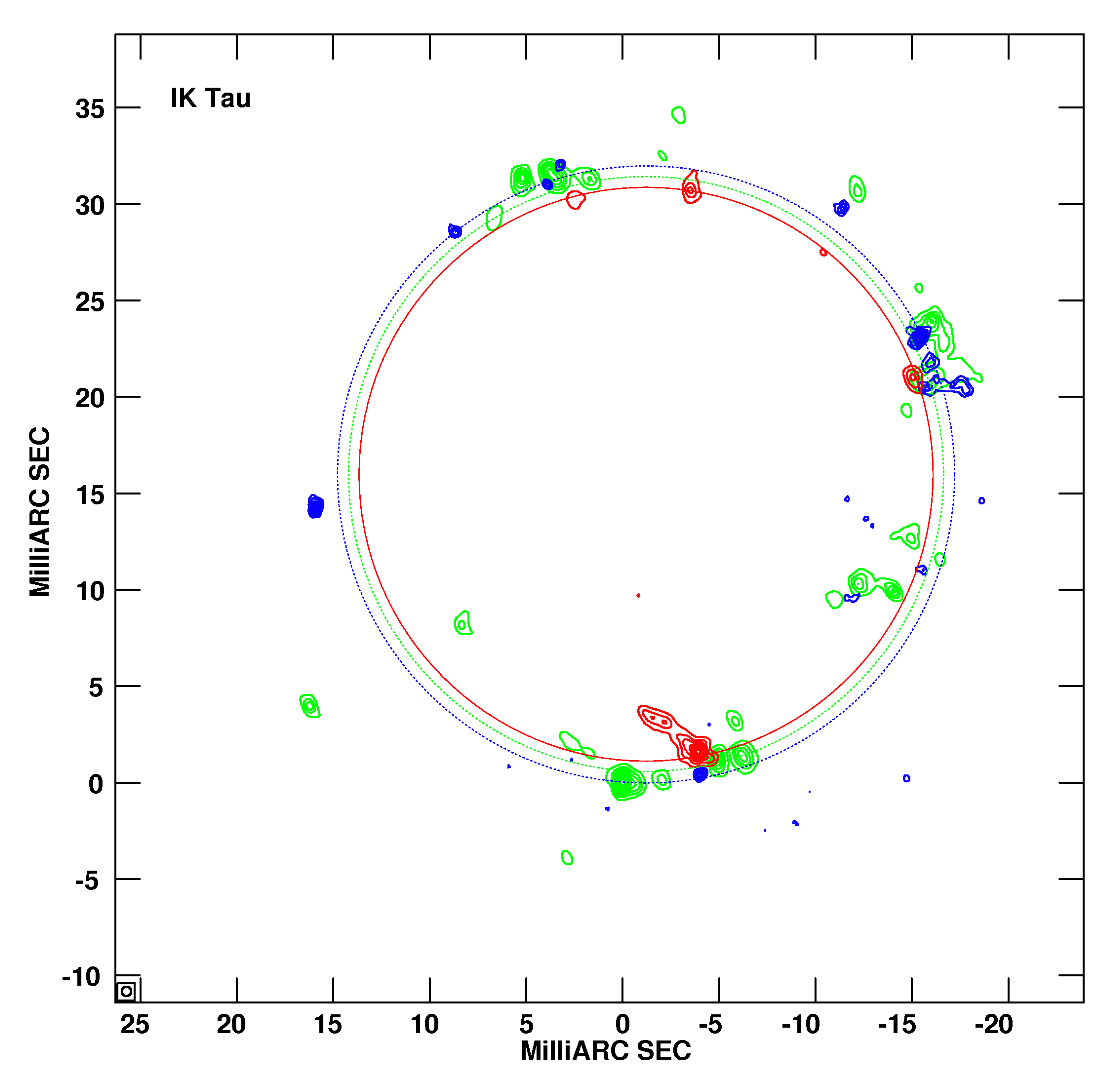}
 \caption{VLBA maps of SiO $J$=1--0 $v$=1 (in blue), $v$=2 (in green),
 and $v$=3 (in red) maser emissions from R~Leo (upper left, November
 13, 2009), TX~Cam (upper right, January 31, 2010), U~Her (lower left,
 April 17, 2011), and IK~Tau (lower right, November 04, 2011). To ease
 the comparison between the three maser lines, using the same color
 code, we plotted the fitting rings obtained with ODRpack for each
 maser transition (see Table \ref{table-fit}).}
   \label{fig-map}
\end{center}
\end{figure*}

We have used the NRAO\footnote {The National Radio Astronomy
Observatory is a facility of the National Science Foundation operated
under cooperative agreement by Associated Universities, Inc} VLBA to
perform quasi-simultaneous observations of the \juc\ lines of
$^{28}$SiO from the first three vibrationally excited states $v$=1,
$v$=2, $v$=3, respectively at 43.122, 42.820, and 42.519~GHz, in both
left and right circular polarizations and using all available
antennas. During the observations, the telescopes periodically switched
every two hours between two frequency setups to record the three lines
(observing $v$=1, $v$=2 and $v$=2, $v$=3). We observed four AGB stars,
R~Leo on November 13, 2009, TX~Cam on January 31, 2010, U~Her on April
17, 2011, and IK~Tau on November 4, 2011.

The data were correlated in the NRAO correlator at Socorro (New Mexico,
USA), which provides 256 frequency channels in the total 8~MHz
bandwidth. We therefore attained a spectral resolution of about
0.2~\kms within a total velocity coverage of 55~\kms, which is adequate
for accurate line profile analysis and covers the total line width.

The calibration was done using the standard procedures for spectral
line VLBI observations in the Astronomical Image Processing System
(AIPS) package. The amplitude was calibrated using the template spectra
method. Bandpass corrections, single-band delay corrections, and phase
errors were corrected by deriving the single-band delay corrections
from continuum calibrators. Corrections of fringe-rates were measured
by selecting a bright and point-like channel emission and then applied
to all the channels.  Note that our observations were done using the
standard line observing mode without including phase referencing
technique, and for the final phase calibration, we have used a maser
component as reference; all positions in the resulting maps are thus
relative to the location of this spot. Therefore, the absolute
positions are lost and only relative positions to the reference channel
are obtained. The maps were produced using the CLEAN deconvolution
algorithm.

In Fig.\ \ref{fig-map}, we show final maps of the brightness
distribution (both polarizations averaged) of $^{28}$SiO $v$=1, $v$=2
and $v$=3, \juc\ (respectively drawn in blue, green, and red), obtained
in R~Leo, TX~Cam, U~Her and IK~Tau.  Each observation run lasted 8
hours, which allows us to reach a rms noise in all the maps of about
5~mJy per channel. All maps were produced with a natural beam
resolution of 0.5~mas.

\section{Results}

\begin{table}[h]
\centering
\small
\begin{tabular}{l|c|c|c|c|c|c}
\hline
\hline
{\bf Source}&\multicolumn{2}{c|}{$v$=1}&\multicolumn{2}{c|}{$v$=2}&\multicolumn{2}{c}{$v$=3}\\
            & Radius & Width& Radius & Width& Radius & Width\\
            &(mas)&(mas)&(mas)&(mas)&(mas)&(mas)\\
\hline
R~Leo  &30.14& 0.43 &28.77 &1.80 &indet.&indet.\\
TX~Cam &16.05& 1.16 &15.63 &1.50 &16.07&1.14\\
U~Her  &10.82& 1.19 &10.43 &1.66 &10.9&0.46\\
IK~Tau &16.02& 0.42 &15.42 &1.57 &14.9 &0.38\\ 
\hline
\end{tabular}
 \caption[]{Summary of the mean radius and width (in mas) obtained for
 the three transitions $^{28}$SiO $v$=1, $v$=2, $v$=3, \juc\ by fitting
 a ring-like brightness distribution to the observed emission using the
 regression package ODRpack.}
\label{table-fit}
\end{table}
\noindent

We have obtained quasi-simultaneous maps of the three 43~GHz maser
lines of SiO, $v$=1, $v$=2 and $v$=3 \juc, as seen in Fig.\
\ref{fig-map}. These are the first VLBA maps ever of the $v$=3 \juc\
maser.  Using VERA, \cite{imai12} observed the $v$=2, $v$=3 \juc\ maser
emissions in the AGB stars, WX~Psc and W~Hya, as seen in \cite{imai10};
a hint of the typical ring-like distribution for the two masers can be
seen in their maps but with a much less clear pattern than in our data.

As the absolute position of each map was lost, the alignment of the
maps presented in Fig.\ \ref{fig-map} was obtained by assuming that the
centroid of all three distributions is coincident (and coincident with
the center of the star) and also by considering the similitude in the
velocity and distribution of the spots. In any case, the relative
positioning of the different transitions seems very reliable in view of
the quite complete ring structures found for all three transitions, and
our conclusions are not affected if other reasonable alignments are
adopted.

All maps show typical clumpy emission with the spots arranged in a well
defined ring distribution.  Using a software for weighted Orthogonal
Distance Regression (ODRpack\footnote{ODRpack is a collection of
Fortran subprograms that can be freely downloaded from
http://www.netlib.org/odrpack/readme}) and taking only the components
with SNR$\ge$5 into account, we estimate the mean radius and width of
the masing emission for all three transitions for our four sources (see
Table \ref{table-fit}). For $v$=3 in R~Leo, it was not possible to fit
any circle \citep[see][for more details on the fitting
process]{soria05} due to the low number of maser clumps and their
weaknesses.  The radius is almost the same for the three
transitions. As already known, the $v$=2 \juc\ maser spots are found to
be placed slightly inward than the $v$=1 emission, but with a very
small difference that is not always detectable (as in the case of
IK~Tau). The $v$=3 \juc\ line emission is placed coincident or slightly
inward (one clear case is the IK~Tau maps) compared to the other two
lines, again with a very small difference. The spots of all three lines
are forming groups, which can in general be identified for the three
lines. However, the individual spots are almost never coincident; even
if we allow some relative shift between the different transitions, it
is impossible to obtain any significant superposition of the spots
(This result was already known for the $v$=1, $v$=2 lines, as seen in
Sect.\ 1).

The most striking result is the very similar ring-like general
distribution of the three lines. This structure is quite different from
that found in $v$=1 \jdu\ maps at 86~GHz, which are also ring-like but
significantly larger. \cite{soria07} found a ring diameter in R~Leo of
67~mas of $v$=1 \jdu\ (They found diameters of 58 and 52~mas for \juc\
$v$=1 and $v$=2, respectively.). In IRC\,+10011, \cite{soria05} found a
ring diameter of about 31.7~mas (and of 23.2 and 21.1~mas for \juc\
$v$=1 and $v$=2, respectively) for $v$=1 \jdu. This result is
particularly difficult to understand if no {\em exotic} mechanism like
line overlap is invoked (Sect.\ 1), since the \juc\ and \jdu\ $v$=1
transitions require practically the same excitation conditions (They
are placed at about 1780 K from the ground and separated by about
4~K.), while levels for \juc\ $v$=2 are 1780~K more energetic than for
the $v$=1 and \juc\ $v$=3, which still requires an additional 1780~K.

 On the other hand, maps of $v$=1 \juc\ and \jdu\ in the S-type star
$\chi$ Cyg show relatively similar distributions \citep{soria04}; we
note that the chemistry of S-type stars is expected to be different
than for the usual O-rich stars with a lower H$_2$O abundance in
particular \citep[see][]{alco13}, and, therefore, less efficient
overlap effects. Remarkably, the $v$=2 \jdu\ and \juc\ maps in this
source are significantly more compact \citep{soria04,cot10}, but the
maps show just a few spots in general and the results are not
conclusive.

%
\begin{figure*}[h]
\begin{center}
   \includegraphics[angle=0,width=0.65\textwidth]{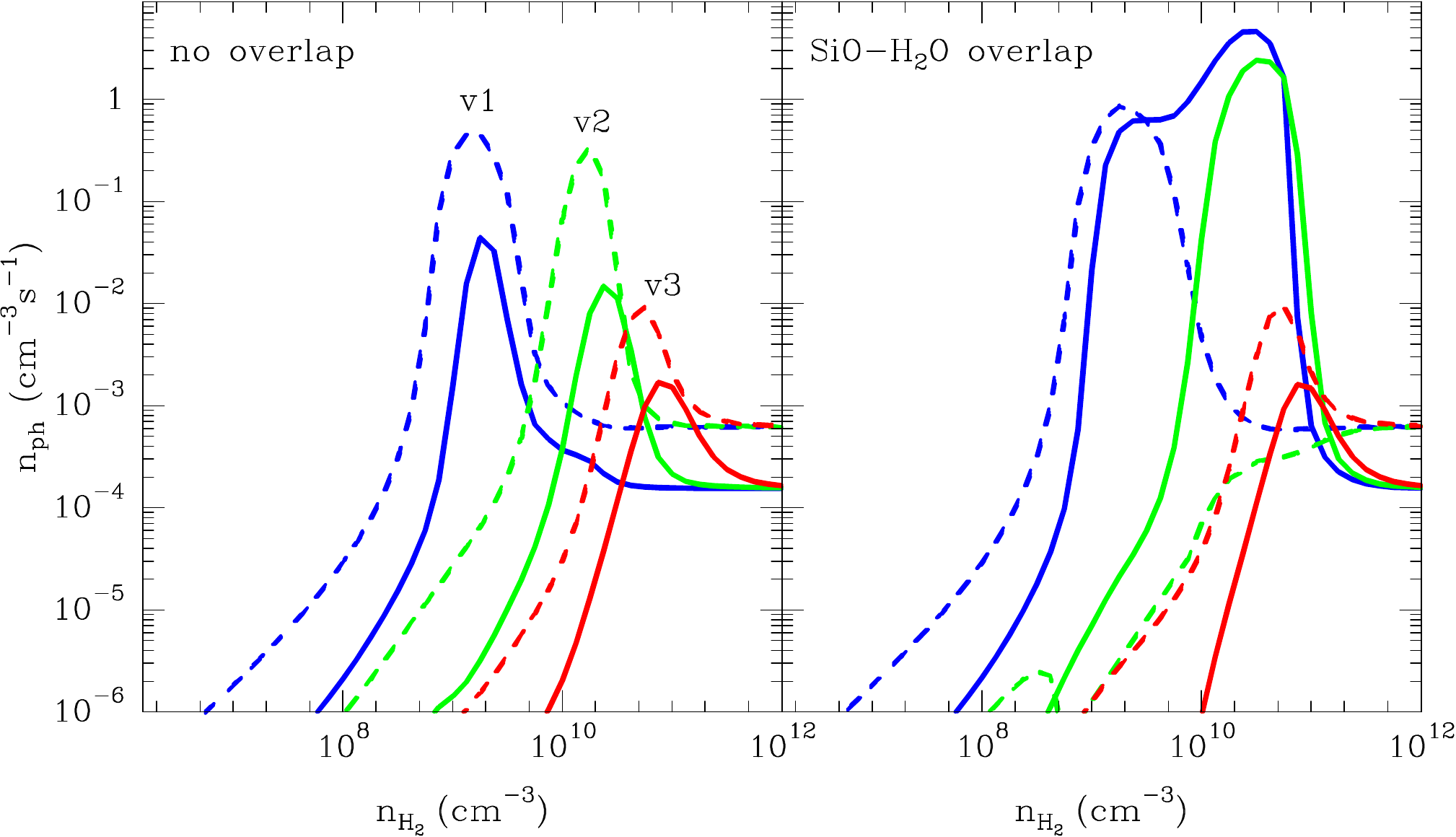}
 \caption{Effects of the SiO-H$_2$O line overlap on the excitation of
 the SiO maser emission for the three first vibrationally excited
 levels, $v$=1, $v$=2, $v$=3 (blue, green and red respectively) \juc\
 (solid lines) and \jdu\ (dashed lines). The number of emitted photons
 (s$^{-1}$\,cm$^{-3}$) in the maser transitions as a function of H$_2 $
 density is shown. Left panel: model calculations that do not include
 the effects of the line overlap. Right panel: model results including
 the line overlap. }
 \label{fig-model}
\end{center}
\end{figure*}

\subsection{Calculations of the SiO $v$=1, $v$=2, and $v$=3 maser pumping in
  the presence of line overlap}

We have developed a theoretical model that calculates the emissivity
and excitation of the SiO maser lines in the circumstellar envelope of
an AGB star. Our code is similar to the one presented by \cite{soria04}
but includes the SiO rotational transitions in the $v$=3 vibrational
state.  In particular, the excitation of SiO is calculated under the
$LVG$ approximation, which approximately takes the maser effects into
account even in the case of saturated masers. This theoretical approach
can be used when there is a large velocity gradient in the masing
region, and therefore, points that are separated a certain distance do
not interact radiatively, resulting in a considerable simplification of
the calculations.  The inversion of the level populations in the maser
transitions is due to the photon self-trapping in the $\Delta v$=1
ro-vibrational transitions when these lines are opaque
\citep[see][]{kwa74}.  When this is the case, the main de-excitation
path for SiO molecules in the $v$\,$\geq$\,1 levels is spontaneous
decay via $\Delta v$=1 transitions.  If the opacity is sufficiently
large, the escape probability in the envelope is given by the inverse
of the opacity for the corresponding transition, which is proportional
to the $J$ quantum number of the upper state. The higher the value of
$J$, the more difficult for an SiO molecule to de-excite via $\Delta
v$=1 transitions, which results in a chain of masers along the
corresponding vibrationally excited state.  This inversion mechanism
only operates when the pumping rate is independent of the $J$ value,
which is expected for collisional excitations and for radiative ones
but only when they are optically thin.  Therefore, radiative models
must include optically thin $\Delta v$=1 excitation and optically thick
de-excitation ro-vibrational transitions at the same time.
This a priori contradiction is solved by assuming a large velocity
gradient or a geometry in which the masers arise from a thin spherical
layer of the envelope, which are expected properties in inner
circumstellar shells. If this is satisfied, the radial opacity in the
$\Delta v$=1 transitions is low, whereas the overall opacity (averaged
in all directions) is high. The calculations we show here are performed
under conditions leading to radiatively excited masers, for which the
effects of line overlap are generally more important
\citep[see][]{buj96,soria04}. We recall that the goal of our paper is
to show that line overlap can explain the observed spatial
distributions of the different masing lines, which not a complete
description of the line overlap effects for all possible physical
conditions and the corresponding pumping properties.
Because of the nature of the inversion mechanism, the physical
conditions required by SiO masers within the same vibrational state
(rotational lines; \juc\ at 43 GHz, \jdu\ at 86 GHz, etc...)  are
similar in either radiative or collisional theoretical models.
However, this is not the case for masers in different vibrationally
excited states ($v$=1, $v$=2, $v$=3 etc.), where the conditions must be
naturally different, since it is difficult to have similar opacities
for the $\Delta v$=1 ro-vibrational transitions of levels with
different excitations in a given shell of the envelope.

As mentioned in previous sections, \cite {soria04} showed that the
observed relative location of the SiO masing regions was not reproduced
by any of the radiative or collisional model, but may be explained by
introducing the effects of the line overlaps between ro-vibrational
transitions of $^{28}$SiO and H$_{2}$O.  This alternative mechanism was
first proposed by \cite{olof81} to explain the anomalous weakness of
the $v$=2 \jdu\ SiO maser in O-rich stars and its relatively high
intensity in S-type stars.  In particular, this was explained by
invoking the overlap between the $v$=1,$J$=0--$v$=2,$J$=1 line of
$^{28}$SiO and the $v_2$=0--1 $J_{K_a,K_c}$=$12_{7,5}$--$11_{6,6}$ line
of para-H$_{2}$O, which are two lines with frequencies coincident down
to a fraction equivalent to less than 1~\kms.

This phenomenon was studied in more detail by \cite {buj96}, who found
that, the observed properties of the $v$=2 \jdu\ maser of $^{28}$SiO in
both O-rich and S-type stars could be explained by introducing this
line overlap in the model calculations. The line overlap efficiently
affects the SiO masers because emission dominates absorption in the
H$_2$O line under expected conditions, which leads to a relative
increase of the radiation available at the frequency of both
vibrational transitions and a net increase of the number of photons
absorbed by the $v$=1,$J$=0--$v$=2,$J$=1 SiO line (with respect to the
case in which line overlap is not considered). Accordingly, a
significant relative overpopulation of the $v$=2 $J$=1 level is
produced in O-rich stars and the $v$=2 \jdu\ population inversion is
quenched. As mentioned, this phenomenon also affects the excitation
conditions of other lines, introducing a strong coupling between both
$v$=1, $v$=2, \juc\ masers, since there is significant population
transfer between the $v$=1, $J$=0 and the $v$=2, $J$=1 levels. For more
details on how the line overlaps affect the pumping and level
populations for these transitions, see \cite{buj96, soria04}.

We have performed new calculations, including predictions of the $v$=3,
$J$=1--0 intensity, which are summarized in Fig.\ \ref{fig-model}.  We
represent the number of photons emitted by each maser line per cubic
centimeter, and second, this parameter has been calculated for
different values of the local density. As discussed by \cite{soria04},
the variations of the local density can be interpreted as variations of
the distance to the central star, for simplified assumptions on the
matter distribution in the inner circumstellar shells, the higher
densities appear in general closer to the star. Of course, the exact
equivalence between density and distance to the stellar surface is very
difficult to determine, and here, we only discuss qualitatively the
relative spatial separation of the different maser lines. It is obvious
that the local clumpiness also affects the observed brightness
distribution, which is mostly in the case of these high-amplification
masers that also amplify the local variations of the density. This
yields the well known sharp distribution of the emission in many
independent spots. For the case in which overlap is not considered (see
Fig.\ \ref{fig-model} left) the lines within a given vibrational state
appear under basically the same conditions. On the contrary, lines of
different $v$-states require significantly different conditions, as
expected in view of the very different excitation energies; for
instance, much higher densities (i.e., shorter distances to the star)
are required for $v$=3 lines. Similar conclusions can be obtained from
other published calculations independent of whether the pumping is
dominantly radiative or collisional \citep[e.g.,][]{buj81,loc92}.
However, line overlap effects couple the excitation of the $v$=1 and 2
\juc\ lines, which present their maxima almost under the same
conditions. Moreover, the densities required to pump both lines are now
higher than that in the standard case, which neglects line overlap, and
is quite close to the densities at which $v$=3 \juc\ is intense. These
results are compatible with the observations; three 43~GHz (\juc) lines
appear almost at the same distance from the star and are significantly
closer (higher densities) than the 86~GHz line ($v$=1 \jdu). Even the
small (marginally detected) differences found between the ring
structures of the the three \juc\ lines (i.e.\ the $v$=3 one being
closer than $v$=2 \juc, which itself is slightly closer than $v$=1
\juc) are compatible with the slight difference in the intensity peak
position that is predicted by the model calculations (with $v$=3 \juc\
requiring slightly more density, etc).

\section{Conclusions}

We have observed the VLBA four AGB stars, R Leo, TX Cam, U Her, and IK
Tau, and obtained reliable maps of \juc\ SiO masers in the first three
vibrationally excited states ($v$=1, $v$=2, and $v$=3) toward the four
sources. We find that the spatial brightness distribution of the $v$=3
maser does not show significant differences with respect to those of
the $v$=1 and $v$=2 lines. The $v$=3 maser emission is distributed in a
ring-like pattern and is coincident or slightly inner than those of
$v$=1, $v$=2 \juc. Despite our initial expectation, this agrees with
model predictions and can be easily explained by the range of physical
conditions that give rise to the $v$=1, $v$=2, $v$=3 maser lines, which
are predicted when the effects of the overlaping of two IR lines of SiO
and H$_2$O are taken into account (Sect.\ 3).

When line overlap is not taken into account, the observed distributions
of the SiO maser lines cannot be explained by current models invoking
either collisional or radiative maser pumping. The excitation
conditions are very different for lines within the different
vibrationally excited states ($v$=1, $v$=2, and $v$=3), which are
separated by an energy equivalent to almost 1800 K. On the other hand,
the conditions required to excite the $v$=1 \juc\ and \jdu\ lines are
almost the same, since the energy levels are separated by a few
degrees.  Since standard inversion schemes do not significantly
discriminate the low-$J$ levels, we expect, clearly different spot
distributions for the masers in different $v$-states and very similar
distributions for masers in the same state under both collisional or
radiative pumping mechanisms.

However, the overlap of the above mentioned IR lines significantly
affects the pumping of the $v$=1, $v$=2 \juc: our calculations show
that these two masers are then strongly coupled and require higher
excitation conditions, which is similar to those of the $v$=3 lines
(which are not significantly affected by the line overlap); see details
in Sect.\ 3. Therefore, the $v$=1, $v$=2, $v$=3 \juc\ lines require
quite similar excitation conditions and should appear in practically
the same circumstellar shells due to the overlap effects. However, the
$v$=1 \jdu\ maser, which is practically not affected by the considered
pair of IR lines, requires lower excitation condition, and should
appear in outer shells. Under the physical conditions adopted in our
models we stress that the pumping of the SiO masers is mainly radiative
and, indeed, that the effects of line overlap tend to be more important
in the radiative pumping regime than for the collisional one.  However,
we cannot rule out that collisional models, includes the effects of
line overlap, could also explain the relative spatial distributions of
the different maser lines.

The predictions presented here are, therefore, compatible with the
existing maps of these four maser lines, notably for $J$=1--0
presented in this work and for our previous $v$=1 $J$=2--1 maps
\citep{soria04,soria05,soria07}. Even the small (but systematic)
differences found between the radii at which the three $J$=1--0 maser
emissions appear in the maps are qualitatively compatible with the
model predictions.

\begin{acknowledgements}
 This work has made use of the databases of SIMBAD
(http://simbad.u-strasbg.fr) and the AAVSO (http://http://www.aavso.org) 
\end{acknowledgements}

{}

\begin{table*}[h]
\centering
\large
\vspace*{2mm}
\hspace*{-1cm}
\resizebox{\textwidth}{!}{
\begin{tabular}{|lr|*{12}{c|}}
\hline
{\bf Source}&&{08-12-12}&{09-03-17}&{09-08-13}&{09-11-07}&{10-01-08}&{10-12-06}&{11-03-06}&{11-04-22}&{11-05-18}&{11-06-15}&{11-10-30}&{11-12-09}\\
\hline
IRC+10011 &v1& &{9.6}&{3.6}&{2.9}&{3.0}&{3.1}&{1.9}&{1.6}&{1.5}&{2.1}&{5.3}&{6.1}\\ 
          &v2& & &{4.8}&{4.1}&{4.1}&{6.0}&{3.5}&{3.1}&{3.8}&{4.8}&{7.5}&{8.4}\\ 
          &v3& &{2.6}&{<0.6}&{0.3}&{0.2}&{0.9}&{0.6}&{0.4}&{0.6}&{0.7}&{2.8}&{2.4}\\ 
o Cet     &v1& &{13.8} &{2.0}&{34.8}&{45.6}&{39.3}&{12.9}&{3.1}&{3.1}&{3.5}&{27.7}&{19.3}\\ 
          &v2& & &{5.6}&{14.1}&{33.3}&{30.8}&{7.3}&{3.1}&{2.1}&{2.4}&{26.0}&{17.2}\\
          &v3& &{<0.3}&{<0.6}&{<0.2}&{<0.2}&{<0.2}&{<0.15}&{<0.3}&{<0.2}&{<0.2}&{<0.2}&{<0.2}\\ 
IK Tau    &v1& &{26.8}&{17.5}&{14.2}&{14.4}&{5.5}&{8.1}&{7.0}&{7.6}&{8.1}&{\bf14.7}&{10.7}\\  
          &v2& & &{16.7}&{13.3}&{16.5}&{6.7}&{5.2}&{6.0}&{8.0}&{7.6}&{\bf14.8}&{9.5}\\ 
          &v3& &{2.6}&{<0.4}&{<0.2}&{0.4}&{<0.2}&{0.15}&{0.5}&{0.9}&{1.7}&{\bf2.2}&{0.9}\\ 
U Ori     &v1& &{9.0}&{2.1}&{1.8}&{1.7}&{0.8}&{9.2}&{8.4}&{6.3}&{4.6}&{1.7}&{1.3}\\  
          &v2& & &{2.5}&{1.0}&{2.1}&{5.0}&{11.7}&{9.9}&{6.3}&{4.0}&{1.1}&{0.7}\\ 
          &v3& &{1.3}&{<0.6}&{<0.2}&{<0.2}&{<0.2}&{0.2}&{0.2}&{<0.2}&{<0.15}&{<0.15}&{<0.15}\\ 
TX Cam    &v1&{8.2}&{4.0}&{5.5}&{4.1}&{\bf12.6}&{2.5}&{4.4}&{2.1}&{2.4}&{4.3}&{6.2}&{4.1}\\  
          &v2& & &{1.5}&{7.9}&{\bf15.5}&{4.4}&{6.0}&{5.7}&{6.3}&{6.2}&{10.4}&{6.3}\\ 
          &v3&{0.4}&{<0.15}&{<0.6}&{0.25}&{\bf1.2}&{<0.1}&{<0.1}&{<0.1}&{<0.1}&{0.2}&{0.3}&{0.2}\\ 
V Cam     &v1&{1.5}&&{1.2}&{1.0}&{1.4}&{1.6}&{0.7}&{0.5}&{0.5}&{0.7}&{0.6}&{0.7}\\  
          &v2& & &{1.3}&{1.1}&{1.0}&{2.3}&{1.0}&{0.8}&{0.6}&{0.6}&{1.2}&{1.3}\\ 
          &v3&{<0.2}& &{<0.2}&{<0.1}&{<0.15}&{<0.1}&{<0.1}&{<0.1}&{<0.1}&{<0.1}&{0.3}&{0.6}\\ 
R Leo     &v1& &{2.3}&{30.2}&{\bf40.4}&{28.0}&{8.3}&{20.4}&{25.9}&{25.7}&{24.0}&{19.1}&{19.0}\\  
          &v2& & &{38.6}&{\bf51.9}&{21.8}&{6.9}&{12.1}&{4.0}&{14.0}&{16.0}&{29.4}&{25.3}\\ 
          &v3& &{0.4}&{5.6}&{\bf4.4}&{<0.2}&{<0.1}&{0.7}&{<0.1}&{<0.1}&{0.2}&{<0.2}&{0.4}\\ 
R LMi     &v1&{8.1}&{10.8}&{2.2}&{4.9}&{7.7}&{4.6}&{4.0}&{3.1}&{3.6}&{2.9}&{2.0}&{3.0}\\  
          &v2& & &{4.9}&{3.3}&{5.7}&{1.4}&{2.2}&{1.6}&{2.0}&{2.1}&{6.3}&{9.1}\\ 
          &v3&{0.7}&{1.1}&{<0.4}&{<0.15}&{0.2}&{<0.1}&{<0.1}&{<0.1}&{<0.1}&{<0.1}&{<0.1}&{<0.1}\\ 
U Her     &v1&{8.9}& &{2.8}&{10.3}&{9.3}&{20.3}&{36.2}&{\bf28.6}&{22.2}&{19.5}&{24.1}&{27.1}\\  
          &v2& & &{5.6}&{14.2}&{13.4}&{21.2}&{36.5}&{\bf28.3}&{22.1}&{20.2}&{23.1}&{23.7}\\ 
          &v3&{0.3}& &{<0.4}&{1.8}&{3.4}&{4.7}&{6.0}&{\bf3.3}&{1.2}&{<0.1}&{<0.1}&{<0.15}\\ 
RR Aql    &v1&{1.8}& &{2.8}&{1.4}&{1.6}&{0.9}&{1.5}&{1.6}&{1.6}&{1.5}&{0.8}&{0.7}\\  
          &v2& & &{2.2}&{0.5}&{1.2}&{0.7}&{1.2}&{2.0}&{1.7}&{1.6}&{0.6}&{0.7}\\ 
          &v3&{<0.2}& &{<0.6}&{<0.2}&{0.2}&{<0.2}&{<0.2}&{<0.2}&{<0.2}&{<0.2}&{<0.2}&{<0.15}\\ 
$\mu$ Cep &v1&{<0.6}& &{2.26}&{1.4}&{1.0}&{1.5}&{1.6}&{1.4}&{1.3}&{1.3}&{1.1}&{0.6}\\  
          &v2& & &{2.4}&{1.4}&{1.1}&{1.1}&{0.8}&{0.7}&{0.8}&{1.0}&{0.9}&{0.7}\\ 
          &v3&{<0.2}& &{<0.15}&{0.2}&{0.2}&{<0.1}&{<0.1}&{0.1}&{<0.1}&{<0.1}&{<0.1}&{<0.1}\\ 
R Cas     &v1&{6.8}&{31.2}&{8.7}&{5.4}&{6.0}&{4.4}&{8.8}&{17.0}&{19.6}&{26.9}&{26.2}&{15.9}\\  
          &v2& & &{11.1}&{11.2}&{17.1}&{2.8}&{6.5}&{18.6}&{29.1}&{31.1}&{22.8}&{19.5}\\ 
          &v3&{1.3}&{4.6}&{<0.5}&{<0.15}&{<0.15}&{<0.1}&{<0.1}&{0.2}&{1.0}&{1.8}&{2.6}&{1.2}\\ 
Y Cas     &v1&{2.8}& &{1.7}&{0.7}&{0.9}&{0.9}&{1.1}&{1.2}&{1.7}&{2.7}&{2.9}&{2.1}\\
          &v2& & &{1.3}&{0.9}&{1.1}&{0.8}&{1.0}&{1.8}&{1.9}&{1.7}&{1.9}&{1.2}\\ 
          &v3&{<0.2}& &{<0.4}&{<0.15}&{0.1}&{<0.1}&{0.1}&{0.2}&{0.3}&{0.3}&{0.2}&{<0.1}\\ 
W And     &v1& &{0.6}&{0.6}&{0.2}&{0.4}&{1.3}&{0.2}&{0.14}&{0.2}&{0.3}&{0.6}&{0.4}\\ 
          &v2& & &{0.7}&{0.5}&{0.4}&{1.3}&{0.3}&{<0.1}&{0.2}&{0.2}&{0.5}&{0.5}\\ 
          &v3& &{0.4}&{<0.4}&{<0.15}&{<0.2}&{<0.1}&{<0.1}&{<0.1}&{<0.1}&{<0.1}&{<0.1}&{<0.1}\\ 
R Cnc     &v1& &{2.8}&{3.1}&{3.2}&{1.2}&{2.5}&{0.8}&{1.7}&{1.7}&{1.6}&{2.7}&{1.6}\\  
          &v2& & &{7.2}&{6.1}&{3.3}&{1.1}&{1.0}&{1.3}&{0.8}&{0.6}&{0.9}& {1.2}\\ 
          &v3& &{<0.2}&{<0.5}&{<0.2}&{<0.2}&{<0.1}&{<0.1}&{<0.15}&{<0.1}&{<0.15}&{<0.15}&{<0.2}\\ 
RU Her    &v1&{<0.5}& &{1.7}&{3.9}&{5.6}&{1.6}&{1.6}&{1.9}&{1.5}&{1.7}&{0.3}&{0.3}\\ 
          &v2& & &{4.2}&{6.7}&{5.9}&{1.7}&{1.3}&{1.1}&{0.8}&{0.9}&{0.3}&{0.3}\\ 
          &v3&{<0.2}& &{<0.5}&{1.4}&{1.1}&{<0.4}&{<0.1}&{<0.1}&{<0.1}&{<0.1}&{<0.1}&{<0.1}\\ 
GY Aql    &v1&{<0.1}& &{2.0}&{0.7}&{<0.2}&{6.6}&{3.7}&{2.9}&{2.3}&{2.6}&{5.8}&{10.2}\\  
          &v2& & &{2.5}&{1.7}&{0.9}&{5.5}&{3.1}&{2.3}&{2.5}&{2.4}&{6.1}&{5.5}\\ 
          &v3&{<0.3}& &{<0.5}&{<0.2}&{<0.2}&{<0.15}&{<0.12}&{<0.2}&{<0.2}&{<0.2}&{0.3}&{0.4}\\ 
R Aql     &v1&{l0.2}& &{6.4}&{3.7}&{1.3}&{1.6}&{1.4}&{2.6}&{5.1}&{6.6}&{7.3}&{5.0}\\  
          &v2& & &{11.2}&{2.6}&{0.9}&{3.8}&{5.9}&{10.7}&{7.5}&{6.6}&{5.8}&{4.5}\\ 
          &v3&{<0.2}& &{<0.5}&{<0.2}&{<0.2}&{0.6}&{0.2}&{<0.2}&{<0.15}&{<0.1}&{<0.1}&{<0.15}\\ 
$\chi$ Cyg &v1&{2.1}& &{<0.5}&{0.5}&{2.9}&{3.2}&{5.2}&{5.1}&{5.2}&{4.9}&{1.5}&{1.3}\\  
          &v2& & &{<0.5}&{0.4}&{0.8}&{<0.1}&{0.16}&{<0.15}&{0.2}&{0.2}&{0.5}&{1.5}\\ 
          &v3&{<0.2}& &{<0.5}&{<0.15}&{<0.15}&{<0.2}&{<0.1}&{<0.15}&{<0.1}&{0.1}&{<0.1}&{<0.1}\\ 
\hline
\end{tabular}
}
\caption[]{Summary of our three years of monitoring with the Onsala
 telescope. For each epoch (date format YY-MM-DD), we indicate the peak
 flux or, in the case of no detection, the upper limit. Bold face values
 are the closest observations in time to our VLBA observations.}
\label{table-moni}
\end{table*}
\noindent

\end{document}